\documentclass[12pt]{article}
\usepackage{epsfig}
\usepackage{graphicx}
\usepackage{multicol}
\renewcommand{\theequation}{\thesection.\arabic{equation}}

\def\npb#1#2#3{    {\it Nucl. Phys. }{\bf B #1} (#2) #3}
\def\plb#1#2#3{    {\it Phys. Lett. }{\bf B #1} (#2) #3}

\def\prd#1#2#3{    {\it Phys. Rev. }{\bf D #1} (#2) #3}

\def\prl#1#2#3{    {\it Phys. Rev. Lett. }{\bf #1} (#2) #3}

\def\zpc#1#2#3{    {\it Z. Phys. }{\bf C #1} (#2) #3}

\def\ibid#1#2#3{   {\it ibid. }{\bf #1} (#2) #3}
\def\jhep#1#2#3{   {\it JHEP  }{\bf #1} (#2) #3}

\newcommand{\lsim}{\stackrel{<}{_\sim}}

\newcommand{\ba}{\begin{array}}
\newcommand{\ea}{\end{array}}
\newcommand{\be}{\begin{equation}}
\newcommand{\ee}{\end{equation}}
\newcommand{\bea}{\begin{eqnarray}}
\newcommand{\eea}{\end{eqnarray}}
\newcommand{\beq}{\begin{equation}}
\newcommand{\eeq}{\end{equation}}

\renewcommand\Re{\mbox{Re}}
\renewcommand\a{\alpha}
\newcommand\G{\Gamma}
\newcommand{\cA}{{\cal A}}
\newcommand{\cO}{{\cal O}}
\newcommand{\cL}{{\cal L}}

\newcommand{\no}{\nonumber}
\newcommand{\Cnew}{C^{\rm new}}
\newcommand{\mbMS}{\overline{m}_b}
\newcommand{\mcpole}{m_c}
\newcommand{\mbpole}{m_b}
\newcommand{\mqpole}{m_{b,c}}
\newcommand{\eps}{\epsilon}
\newcommand{\pslash}{{\slash\!\!\! p}}

\begin{document}
 
\thispagestyle{empty}
\begin{flushright}
CERN--TH/2002--161\\
July 2002
\end{flushright}

\vspace*{1.5cm}
\centerline{\Large\bf Forward--Backward Asymmetry in 
$B \rightarrow X_s \ell^+\ell^-$}
\vspace*{0.2cm}
\centerline{\Large\bf at  the NNLL Level}

\vspace*{2cm}
\centerline{{\large\bf  A.~Ghinculov,$^{a,}\footnote{~Work supported by the U.S. Department of Energy (DOE).}$ T.~Hurth,$^{b,}$\footnote{~Heisenberg Fellow.}
 G.~Isidori,$^{b,c}$ Y.-P.~Yao$^{d,1}$ 
}}
\bigskip

\begin{center}
{\it ${}^a$~Department of Physics and Astronomy\\ UCLA, Los Angeles CA 90095-1547, USA}\\
\vspace{0.3cm}
{\it ${}^b$~Theoretical Physics Division, CERN, CH-1211 Geneva 23, Switzerland}\\
\vspace{0.3cm}
{\it ${}^c$~INFN, Laboratori Nazionali di Frascati, I-00044 Frascati, Italy}\\
\vspace{0.3cm}
{\it ${}^d$~Randall Laboratory of Physics\\  University of Michigan, Ann Arbor 
 MI 48109-1120, USA}
\end{center}

\vspace*{1.5cm}

\centerline{\large\bf Abstract}
\begin{quote}
We report the results of a new calculation of soft-gluon corrections 
in $B \rightarrow X_s \ell^+\ell^-$ decays. In particular,
we present the first calculation of bremsstrahlung and corresponding
virtual terms to the lepton 
forward--backward asymmetry, which allows us to 
systematically include all 
contributions to this observable beyond the lowest non-trivial order. 
The new terms  are important, for instance
the position of the zero of the asymmetry
receives corrections of $O(10\%)$. Using a different method,
 we also provide
an independent check of recently published results 
on bremsstrahlung and infrared virtual corrections to the 
dilepton-invariant 
mass distribution. 
\end{quote}
\vfill
\newpage
\pagenumbering{arabic}

\setcounter{equation}{0}
\section{Introduction}
\label{intro}
Flavour-changing neutral-current (FCNC) processes are a 
very useful tool to understand the nature of physics beyond the 
Standard Model (SM). The stringent bounds 
obtained from  $B \rightarrow X_s \gamma$ on 
various non-standard scenarios
(see e.g.~\cite{DGIS,Gambino,Borzumati,Hurth})
are a clear example of the importance of 
theoretically clean FCNC observables
in discriminating new-physics models. 

Generally, inclusive rare decay modes of the $B$ meson 
are theoretically  clean observables. For instance 
the decay width $\G(B \to X_s \gamma)$ 
is well approximated by the partonic decay rate
$\G(b\to s \gamma)$, which can be
analysed in renormalization-group-improved 
perturbation theory. Non-perturbative contributions play
only a subdominant role and can be calculated in 
a model-independent way by using the heavy-quark expansion.
The inclusive $B \rightarrow X_s \ell^+\ell^-$ transition,
which is starting to be accessible at $B$ factories \cite{Bsll_exp}, 
represents a new source of theoretically clean observables, 
complementary to the $B \rightarrow X_s \gamma$ rate.
In particular, kinematic observables such as the invariant 
dilepton mass spectrum and the forward--backward (FB) asymmetry
in \mbox{$B \rightarrow X_s \ell^+\ell^-$} provide additional clean information 
on short-distance couplings not accessible in $B \rightarrow X_s \gamma$
\cite{AliMannel}.

Non-perturbative corrections to $B \rightarrow X_s \ell^+\ell^-$ 
scaling with $1/m_b^2$ and $1/m_c^2$ can be calculated 
quite analogously to those entering the $B \rightarrow X_s \gamma$
rate \cite{Falk,Alineu,Rey,BI}. Here the $c\bar{c}$ resonances
represent a more serious problem since, for specific 
values of the dilepton invariant mass, $c\bar{c}$ states 
can be produced on shell. However,  
these resonances can be removed by appropriate  kinematic cuts
in the invariant mass spectrum. In the {\em perturbative window}, namely 
when $m_{\ell^+\ell^-}\lsim~m_b/2$,
theoretical predictions for the invariant mass spectrum
are dominated by the purely perturbative contributions, 
and a theoretical precision comparable with  the one reached  
in the decay $B \rightarrow X_s \gamma$ is possible. 
The $B$ factories will soon provide statistics and resolution
needed for the measurements of $B \rightarrow X_s \ell^+\ell^-$
kinematic distributions. Precise theoretical estimates of 
the SM expectations are therefore needed in order to perform new 
significant tests of flavour physics.

In this paper we complete one important step of this program:
we present a new calculation of QCD-bremsstrahlung
and the corresponding virtual corrections in $B \rightarrow X_s \ell^+\ell^-$.
This effect has already been evaluated in Refs.~\cite{Asa1,Asa2} 
for the dilepton invariant-mass spectrum. Here we perform 
the calculation using a different technique, namely using 
a full dimensional regularization of infrared divergences 
(both soft and collinear ones). 
We anticipate that our results agree with 
those of \cite{Asa1,Asa2} for the dilepton invariant-mass distribution.
Moreover, our technique allows us obtain the first computation 
of the soft-gluon (and corresponding virtual) corrections 
in the FB asymmetry: 
the missing ingredient needed for a systematic evaluation 
of this observable beyond the lowest non-trivial order.
Our phenomenological analysis shows that 
these new contributions are rather important, 
for instance the position of the zero of the asymmetry 
receives corrections of $O(10\%)$. 

The paper is organized as follows: in Section 2 we set up
the framework of the calculation; we briefly review 
the status of the various ingredients needed for a systematic 
analysis of the partonic process $b \rightarrow s (g) \ell^+\ell^-$ 
in perturbation theory. In Section 3 we present the basic 
expressions needed to describe the spectrum and the FB asymmetry
of $B \rightarrow X_s \ell^+\ell^-$ in the so-called
next-to-next-to-leading logarithmic (NNLL) approximation. 
The details of the calculation, 
including a definition of the four-particle phase space
and a discussion on the regularization scheme, which avoids any  
ambiguity due to the definition $\gamma_5$, are presented 
in Section~4. Section~5 contains a brief phenomenological 
analysis of our results, mainly focused on the determination 
of the zero of the FB asymmetry, and Section 6 
a short summary. 

\setcounter{equation}{0}
\section{Theoretical framework}
Within inclusive $B$ decay modes such as $B \rightarrow X_s \gamma$ or 
$B \rightarrow X_s \ell^+\ell^-$, short-distance QCD corrections 
lead to a sizeable modification of the pure electroweak
contribution, generating large logarithms 
of the form $\alpha_s^n(m_b)\,\log^m(m_b/M_{\rm heavy})$,
where $M_{\rm heavy}=O(M_W)$ and $m \le n$ (with $n=0,1,2,...$).
A suitable framework to achieve the necessary resummations 
of these large logs is the construction of an effective low-energy 
theory with five quarks, obtained by integrating out the
heavy degrees of freedom. With the help of renormalization-group (RG) 
techniques, one can then resum the series of 
leading logarithms (LL), next-to-leading logarithms (NLL), and so on:
\be
\alpha_s^n(m_b) \,  \log^n(m_b/M)\quad [\mbox{LL}], \qquad  
\a_s^{n+1}(m_b) \, \log^n (m_b/M)\quad [\mbox{NLL}]~, \ldots
\label{LLseries}
\ee

The effective five-quark low-energy Hamiltonian 
relevant to the partonic process 
$b \rightarrow s \ell^+\ell^-$ can be written as 
\begin{equation}
{\cal H}_{eff} = - \frac{4 G_{F}}{\sqrt{2}} V_{ts}^*V_{tb} \, 
\sum  {C_{i}(\mu, M_{\rm heavy})}\,\, \, {\cal O}_i(\mu) \, ,
\label{eq:effH}
\end{equation}
where 
\begin{equation}
\begin{array}{rlrl}
{\cal O}_{1} ~= &\!\!
(\bar{s} \gamma_\mu T^a P_L c)\,  (\bar{c} \gamma^\mu T_a P_L b)\,, & 
{\cal O}_{2} ~= &\!\!
(\bar{s} \gamma_\mu P_L c)\,  (\bar{c} \gamma^\mu P_L b)\,,  \\[1.2ex]
{\cal O}_{3} ~= &\!\!
(\bar{s} \gamma_\mu P_L b) \sum_q (\bar{q} \gamma^\mu q)\,,   &  
{\cal O}_{4} ~= &\!\!        
(\bar{s} \gamma_\mu T^a P_L b) \sum_q (\bar{q} \gamma^\mu T_a q)\,, \\[1.2ex]
{\cal O}_{5} ~= &\!\! 
(\bar{s} \gamma_\mu \gamma_\nu \gamma_\rho P_L b) 
 \sum_q (\bar{q} \gamma^\mu \gamma^\nu \gamma^\rho q)\,, &
{\cal O}_{6} ~= &\!\! 
(\bar{s} \gamma_\mu \gamma_\nu \gamma_\rho T^a P_L b) 
 \sum_q (\bar{q} \gamma^\mu \gamma^\nu \gamma^\rho T_a q)\,, \\[2.0ex]
\widetilde{{\cal O}}_{7}   ~= &\!\!     
  \displaystyle{\frac{e}{16\pi^2}} \, \mbMS(\mu) \,
 (\bar{s} \sigma^{\mu\nu} P_R b) \, F_{\mu\nu}\,,    &
\widetilde{{\cal O}}_{8}   ~= &\!\!  
  \displaystyle{\frac{g_s}{16\pi^2}} \, \mbMS(\mu) \,
 (\bar{s} \sigma^{\mu\nu} T^a P_R b)
     \, G^a_{\mu\nu}\,,        \\[2.0ex]                       
\widetilde{{\cal O}}_{9}   ~= &\!\!         
  \displaystyle{\frac{e^2}{16\pi^2}} \,
 (\bar{s} \gamma_\mu  P_L b)\, (\bar{\ell} \gamma^\mu \ell)\,, &
\widetilde{{\cal O}}_{10}  ~= &\!\! 
  \displaystyle{\frac{e^2}{16\pi^2}} \,
 (\bar{s} \gamma_\mu P_L b)\,  (\bar{\ell} \gamma^\mu \gamma_5 \ell)
\end{array}
\label{heffll}                                        
\end{equation}
define the complete set of relevant dimension-6 
operators and $C_{i}(\mu, M_{\rm heavy})$ 
the corresponding Wilson coefficients.

Within this framework, QCD corrections are twofold: 
corrections related to the Wilson coefficients,
and those related to the matrix elements of the various operators,
both evaluated at the low-energy scale $\mu \approx  m_b$.
As the heavy fields are integrated out, the top-quark-,
$W$-, and $Z$-mass dependence is contained in the 
initial conditions of the Wilson coefficients, 
determined by a matching procedure between full and 
effective theory at the high scale (Step~1). By means of RG equations, 
the $C_{i}(\mu, M_{\rm heavy})$ are then evolved at the low
scale (Step~2). Finally, the QCD corrections to the matrix 
elements of the operators are evaluated at the low scale (Step~3). 

Compared with the effective Hamiltonian relevant to 
$b \to s \gamma$, Eq.~(\ref{heffll}) 
contains the $O(\alpha_{\rm em})$
additional operators $\widetilde{{\cal O}}_{9}$ and  $\widetilde{{\cal O}}_{10}$.
Moreover, it  turns out that 
the first large logarithm 
of the form $\log(m_b/M_W)$ arises already without 
gluons, because the operator ${\cal O}_2$ mixes into $\widetilde{{\cal O}}_9$ 
at one loop. This possibility, which has no equivalent  
in the $b \to s \gamma$ case, leads
to the following ordering of contributions to the decay amplitude 
\bea
&& \left[ \alpha_{\rm em}\, \log(m_b/M) \right]\,
   \a_s^{n}(m_b)\, \log^n(m_b/M) \qquad [\mbox{LL}]~, \no \\  
&& \left[ \alpha_{\rm em}\, \log(m_b/M) \right]\,
   \a_s^{n+1}(m_b)\, \log^n (m_b/M) \quad [\mbox{NLL}]~, \ldots
\eea 
Technically, to perform the resummation, it is convenient
to transform these series into the standard form (\ref{LLseries}).
This can be achieved by redefining magnetic, chromomagnetic and lepton-pair operators 
as follows~\cite{MM,BurasMuenz}:
\begin{equation}
\label{reshuffle}
{\cal O}_i = \frac{16\pi^2}{g_s^2} \widetilde{{\cal O}}_i~, 
\quad C_i = \frac{g_s^2}{(4\pi)^2} \widetilde{C}_i~, \quad \quad (i=7,...,10). 
\end{equation}

This redefinition
enables one to proceed in the standard way, or as in 
$b \to s \gamma$, in the three 
calculational steps discussed above \cite{MM,BurasMuenz}.
At the high scale, the Wilson coefficients can be computed at a
given order in perturbation theory and expanded in powers of 
$\alpha_s$:
\begin{equation} 
\label{Wilson}
C_i = C^{(0)}_i + \frac{\alpha_s}{(4 \pi)} C^{(1)}_i  
+ \frac{\alpha_s^2}{(4\pi)^2} C^{(2)}_i  + ...
\end{equation}
Obviously, the Wilson coefficients of the new operators 
${\cal O}_{7-10}$ at the high scale start at order $\alpha_s$ only. 
Then the anomalous-dimension matrix has the canonical 
expansion in $\alpha_s$ and starts with a term
proportional to $\alpha_s$:
\begin{equation}
\label{dimmatrix}
{\gamma}
= \frac{\alpha_s}{4\pi} {\gamma}^{(0)}
+ \frac{\alpha_s^2}{(4\pi)^2} {\gamma}^{(1)}
+ \frac{\alpha_s^3}{(4\pi)^3} {\gamma}^{(2)} + ... 
\end{equation}
In particular,  after the reshufflings in (\ref{reshuffle})
the one-loop mixing of the operator ${\cal O}_2$ 
with  ${\cal O}_9$ appears 
formally at order $\alpha_s$.

The last of the three steps, however, 
requires some care: among the operators with a non-vanishing 
tree-level matrix element, only  ${\cal O}_9$ has a non-vanishing 
coefficient at the LL level. Therefore, at this level, only the tree-level 
matrix element of this operator ($\langle {\cal O}_9 \rangle$)
has to be included. At NLL accuracy
the QCD one-loop contributions to 
$\langle {\cal O}_9 \rangle$, the tree-level contributions to 
$\langle {\cal O}_7 \rangle$ 
and $\langle {\cal O}_{10}\rangle$, and the electroweak 
one-loop matrix elements of the four-quark operators have to be 
calculated. Finally, at NNLL precision, one should  in principle 
take into account the QCD two-loop corrections 
to $\langle {\cal O}_9 \rangle$, the QCD one-loop corrections to
$\langle {\cal O}_7\rangle $ and  $\langle{\cal O}_{10}\rangle$, 
and the QCD corrections to the electroweak one-loop matrix elements 
of the four-quark operators. 
                       
\medskip 

Let us briefly review the present status 
of these perturbative contributions to decay rate and 
FB asymmetry of $B \to X_s \ell^+\ell^-$: the 
complete NLL contributions to the decay amplitude
can be found in \cite{MM,BurasMuenz}. 
Since the LL contribution to the rate turns out to be numerically 
rather small, NLL terms represent an $O(1)$ correction 
to this observable. On the other hand, since a non-vanishing 
FB asymmetry is generated by the interference of vector  ($\sim {\cal O}_{7,9}$) 
and axial-vector ($\sim {\cal O}_{10}$) leptonic currents,
the LL amplitude leads to a vanishing result and NLL terms represent 
the lowest non-trivial contribution to this observable.

For these reasons, a computation of NNLL terms
in $B \to X_s \ell^+\ell^-$ is needed 
if we aim at the same numerical accuracy as achieved 
by the NLL analysis of $B \to X_s \gamma$ \cite{Mikolaj}.
Large parts of the latter can be taken over 
and used in the NNLL calculation of
$B \to X_s \ell^+\ell^-$. However, this is not the full story. 
In particular, the bremsstrahlung corrections presented in this paper
for the first time
are a crucial ingredient, necessary for a complete evaluation 
of the FB asymmetry to NNLL precision.
And in the case of the differential rate,
the full NNLL enterprise is still not complete:

\begin{description} 
\item{[Step 1]} The full computation of initial conditions 
to NNLL precision has been presented in Ref.~\cite{MisiakBobeth}.
The authors did the two-loop matching 
for all the operators relevant to $b \to s \ell^+\ell^-$
(including a confirmation of the $b \to s \gamma$ NLL matching 
results of \cite{Adel,GH}).
The inclusion of this  NNLL contribution removes the large
matching scale uncertainty (around $16 \%$) of the 
NLL calculation of the $b \to s \ell^+\ell^-$ 
decay rate.

\item{[Step 2]} Thanks to the reshufflings of the LL series,
most of the NNLL contributions to the anomalous-dimension matrix
can be derived from the NLL analysis of $b \to s \gamma$.
In particular the three-loop mixing of the four-quark operators 
${\cal O}_{1-6}$ into ${\cal O}_7$ and 
${\cal O}_8$ can be taken over from Ref.~\cite{Mikolaj}. The only 
missing piece for a full NNLL analysis of  the $b \rightarrow s \ell^+\ell^-$
decay rate is the three-loop mixing of the four-quark operators into ${\cal O}_9$.
In \cite{MisiakBobeth} an estimate was made, which suggests that the numerical 
influence of these missing NNLL contributions on the branching 
ratio of $b \rightarrow s \ell^+\ell^-$ is small. 
Interestingly, since the FB asymmetry has no contributions 
proportional to $|\langle \cO_9 \rangle|^2$, 
this missing term is not needed for a NNLL analysis of this observable.

\item{[Step 3]} Within the $B \rightarrow X_s \gamma$ calculation
at NLL, the two-loop matrix elements of the four-quark operator ${\cal O}_2$
for an on-shell photon were calculated in \cite{GHW}
and quite recently confirmed in \cite{Burasnew}. An independent 
numerical check of these results has been performed and will be presented 
in \cite{in_prog}. This calculation 
was extended in \cite{Asa1} to the case of an off-shell photon (for small
squared dilepton mass),
which corresponds to a NNLL contribution relevant to the decay
$B \rightarrow X_s \ell^+\ell^-$. In the dilepton spectrum, 
this calculation reduces the error 
corresponding to the uncertainty of the low-scale dependence 
from $\pm 13\%$ down to $\pm 6.5\%$. This step also includes 
the bremsstrahlung and virtual contributions that are
discussed in the present paper. \\
In principle, a complete NNLL calculation
of the $B \rightarrow X_s \ell^+\ell^-$ rate would require also 
the calculation of the renormalization-group-invariant 
two-loop matrix element of the operator ${\cal O}_9$. 
Similarly to the missing piece of 
the anomalous-dimension matrix, also this (scale-independent)
contribution does not enter the FB asymmetry at NNLL accuracy.

\end{description}

\setcounter{equation}{0}
\section{Basic expressions}

We normalize all the observables by the semileptonic
decay rate in order to reduce the uncertainties due 
to bottom quark mass and CKM angles: 
\begin{equation}
\Gamma[ b \to X_c e \bar{\nu}_e] = 
\frac{ G_F^2 \mbpole^5 }{ 192 \pi^3} |V_{cb}|^2 f(z) \kappa(z)~.
\end{equation}
Here $z=\mcpole^2/\mbpole^2$ ($\mqpole$ denote pole quark masses), 
$f(z) = 1 - 8 z + 8 z^3 - z^4 - 12 z^2 \ln z$
is the phase-space factor and
\begin{equation}
\kappa(z) = 1 - \frac{2 \alpha_s(m_b)}{3 \pi} \frac{h(z)}{f(z)}
\label{semi}
\end{equation}
takes into account QCD corrections  
(the function $h(z)$ has been given analytically in \cite{NirNir}
and is quoted in the appendix).
The normalized dilepton invariant mass spectrum is then defined as
\begin{equation}\label{decayamplitude}
R(s)=\frac{\frac{d}{d s}\Gamma(  B\to X_s\ell^+\ell^-)}{
\Gamma( B\to X_ce\bar{\nu})}~,
\end{equation}
where $s=(p_{\ell^+}+p_{\ell^-})^2/\mbpole^2$.
The other important observable is the forward--backward 
lepton asymmetry:
\begin{equation}\label{forwardbackward}
A_{\rm FB}(s)=\frac{1}{\Gamma( B\to X_ce\bar{\nu})}
  \int_{-1}^1 d\cos\theta_\ell ~
 \frac{d^2 \Gamma( B\to X_s \ell^+\ell^-)}{d s ~ d\cos\theta_\ell}
\mbox{sgn}(\cos\theta_\ell)~,
\end{equation}
where $\theta_\ell$ is the angle between $\ell^+$ and $B$ momenta 
in the dilepton centre-of-mass frame. It was  shown in \cite{Alineu}
that $A_{\rm FB}(s)$ is identical to the energy asymmetry introduced in 
\cite{WylerMisiakCho}.
 
Following closely --- but not exactly --- the notation of Refs.~\cite{Asa1,Asa2},
we present here some useful formulae that allow us to systematically 
take into account corrections beyond the NLL level for the partonic
 contributions to these two observables:
\bea
 R(s) = \frac{\alpha_{\rm em}^2}{4\pi^2}
\left|\frac{V_{tb}^* V_{ts}}{V_{cb}}\right|^2  && \!\!\!\!\!\!\!\! 
\frac{(1-s)^2}{f(z)\kappa(z)}    \left\{
   4 \left(1+\frac{2}{s}\right) |\Cnew_7(s)|^2   
   \left( 1+ \frac{\alpha_s}{\pi} \tau_{77}(s) \right) \right. \nonumber \\
&&  +(1+2s) \left[|\Cnew_9(s)|^2+|\Cnew_{10}(s)|^2\right] 
   \left(1+ \frac{\alpha_s}{\pi} \tau_{99}(s) \right)  \nonumber \\
&& \left. + 12\, \Re\left[ \Cnew_7(s)  \Cnew_9(s)^* \right] \left(1+  
   \frac{\alpha_s}{\pi} \tau_{79}(s) \right) + \frac{\alpha_s}{\pi}\delta_R(s) \right\}~,
   \qquad \label{NNLLDIMS} \\ 
\nonumber \\
A_{\rm FB}(s) = - \frac{3\alpha_{\rm em}^2}{4\pi^2}
 \left|\frac{V_{tb}^* V_{ts}}{V_{cb}}\right|^2  && \!\!\!\!\!\!\!\! 
 \frac{ (1-s)^2}{f(z)\kappa(z)}   \left\{  
   s \, \Re\left[ \Cnew_{10}(s)^* \Cnew_9(s) \right] 
  \left(1 +  \frac{\alpha_s}{\pi} \tau_{910}(s)\right) \right.  \nonumber\\
&& \left. + 2 \, \Re\left[ \Cnew_{10}(s)^* \Cnew_7(s) \right]
  \left( 1 + \frac{\alpha_s}{\pi}\tau_{710}(s) \right) + \frac{\alpha_s}{\pi}\delta_{\rm FB}(s)
  \right\}~.
\label{NNLLAFB}
\eea
With respect to Refs.~\cite{Asa1,Asa2}, we have introduced 
a new set of effective coefficients, defined as 
\bea
  \Cnew_7(s) &=&  \left(1+\frac{\alpha_s}{\pi}\sigma_7 (s)\right) 
       \widetilde C_7^{\rm eff} 
        -\frac{\alpha_s}{4\,\pi} \left[ C_1^{(0)} F_1^{(7)}(s)+
        C_2^{(0)} F_2^{(7)}(s) 
       + \widetilde C^{\rm eff(0)}_8 F_8^{(7)}(s) \right]  \nonumber \\
  \Cnew_9(s)  &=&  \left(1+\frac{\alpha_s}{\pi} \sigma_9 (s) \right) 
        \widetilde C_9^{\rm eff}(s) 
       -\frac{\alpha_s}{4\,\pi} \left[ C_1^{(0)} F_1^{(9)}(s) 
        + C_2^{(0)} F_2^{(9)}(s)+ \widetilde C_8^{\rm eff (0)} F_8^{(9)}(s) \right]
    \nonumber \\
  \Cnew_{10}(s)  &=& \left( 1+\frac{\alpha_s}{\pi} 
        \sigma_{9} (s) \right) \widetilde C^{\rm eff}_{10}~.
\label{effmod}
\eea
The $\Cnew_i$  have the advantage of encoding all dominant 
matrix-element corrections, which leads to an explicit $s$ dependence
in all of them.

As we shall 
illustrate in detail in the next section, the terms $\sigma_i(s)$ 
take into account {\em universal} $O(\alpha_s)$ bremsstrahlung 
and the corresponding infrared (IR) virtual  corrections proportional 
to the tree-level matrix 
elements of ${\cal O}_{7-10}$. The remaining (finite) non-universal 
bremsstrahlung and IR virtual corrections are encoded in rate and FB asymmetry 
through  $\tau_i(s)$ and $\delta_{R,{\rm FB}}(s)$. Analytic 
and numerical results for the $\tau_i(s)$ and the 
 $\sigma_i(s)$ will be presented later on.
Here we simply note that the  $\sigma_i(s)$ are defined in order to 
take into account the truly soft component of the radiation, 
which diverges at the $s\to 1$ boundary of the phase space. 
With such definition of the $\sigma_i(s)$,
the additional finite terms $\tau_i(s)$ are rather small
all over the phase space and particularly
for large values of $s$ ($|\tau_i(s)|< 0.5$ for $s>0.3$).
Substantially smaller are the   
bremsstrahlung corrections not related to ${\cal O}_{7-10}\otimes {\cal O}_{7-10}$,
which are encoded in $\delta_{R,{\rm FB}}(s)$. A complete evaluation of  
$\delta_{R}(s)$ can be found in~\cite{Asa2}, where     
this effect is shown to be at the $O(1\%)$ level.

The other explicit $O(\alpha_s)$ terms in (\ref{effmod}) are 
due to virtual corrections that are infrared-safe. In particular, 
the two-loop functions 
$F_{1,2}^{(7,9)}$ and the one-loop functions $F_8^{(7),(9)}$ correspond 
to virtual corrections to ${\cal O}_{1,2}$  and ${\cal O}_8$, respectively. 
These functions have been computed in Ref.~\cite{Asa1} for small $s$.
The coefficients $\widetilde C^{\rm eff}_{7-10}$, including the 
one-loop matrix-element contributions of ${\cal O}_{1-6}$ 
are defined in close analogy with Ref.~\cite{BurasMuenz} and are reported 
in the appendix as a function of the true Wilson coefficients $C_i$. 
Finally, explicit expressions for the latter, evolved down at the 
low-energy scale, can be found in \cite{MisiakBobeth}.

\medskip 

By means of expressions (\ref{NNLLDIMS}) and (\ref{NNLLAFB}), 
we can more easily discuss the organization of the perturbative 
expansion for the two observables. According to the arguments 
in Section~2, the formally leading terms 
are obtained by setting $\Cnew_7=\Cnew_{10}=0$, neglecting 
$\tau_i$ and $\delta_i$, and identifying $\Cnew_9$ with 
the leading term of $\widetilde{C}_9^{\rm eff}$ [formally $O(1/\alpha_s)$].
At this level $A_{\rm FB}(s)$ is clearly vanishing. 
At the NNL level, when $A_{\rm FB}(s)$ receives its first 
non-vanishing contribution, we should retain the interference 
of the $O(1/\alpha_s)$ term in $\Cnew_9$ with the leading $O(1)$
terms in $\Cnew_{7,10}$, as well as the
subleading corrections in $|\Cnew_9|^2$. 
Within this approach, the missing 
ingredients for a NNLL analysis of $A_{\rm FB}(s)$
are only $\sigma_9$ and $\tau_{910}$. 

As anticipated, the standard LL expansion is numerically not well 
justified, since the formally-leading $O(1/\alpha_s)$
term in $\Cnew_9$ is much closer to an $O(1)$ term.  
For this reason,  it has been proposed in Ref.~\cite{Asa1} to use 
a different counting rule, where the $O(1/\alpha_s)$
term of $\Cnew_9$ is treated as $O(1)$. We also believe 
that this approach, although it cannot be consistently 
extended at higher orders, is well justified 
at the present status of the calculation. Within this 
approach, the three $\Cnew_i$ and the two observables 
[$R(s)$ and $A_{\rm FB}(s)$] are all  
homogeneous quantities, starting with an $O(1)$ term.
Then all  $\sigma_i$, $\tau_i$ and $\delta_i$ functions are 
required for a next-to-leading order analysis of 
both $R(s)$ and $A_{\rm FB}(s)$. 

\setcounter{equation}{0}
\section{Details of the calculation}
\label{calculation}
In contrast to previous calculations of $B\to X_s\gamma$
and $B\to X_s\ell^+\ell^-$ matrix elements,  
we set $m_s=0$ and, following the approach of Ref.~\cite{Peccei},
we employ dimensional regularization 
to regulate all infrared singularities, including 
collinear divergences arising from the vanishing strange-quark 
mass. 

\begin{figure}
\begin{center}
\epsfig{file=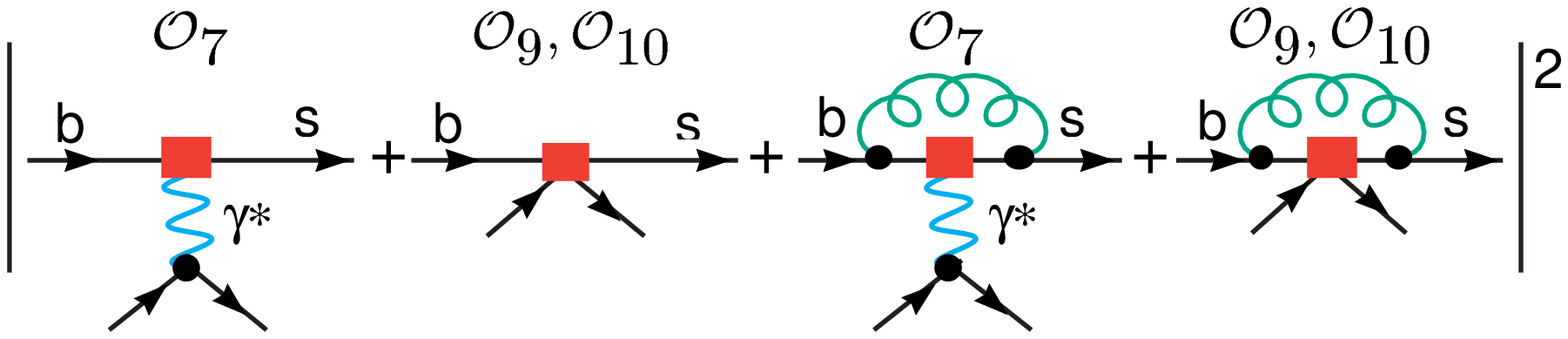,width=14.3cm}
\epsfig{file=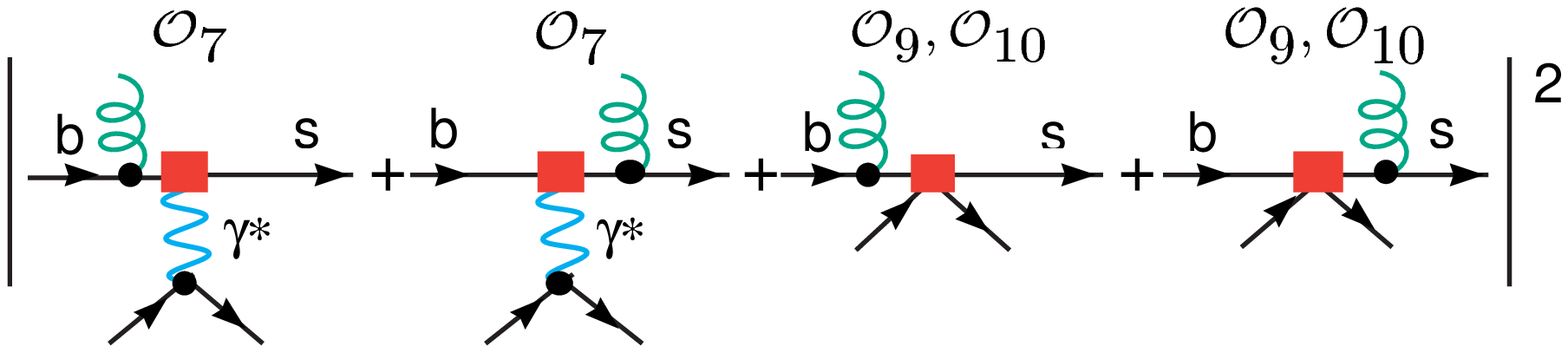,width=14.3cm}
\end{center}
\caption{Virtual (up) and real (down) QCD corrections generating the 
terms $\tau_i$ and $\sigma_i$ in Eqs.~(\ref{NNLLDIMS})--(\ref{effmod}).
The boxes denote the insertion of $\cL_{\rm eff}$ in (\ref{eq:effL}).}
\label{bsll03}
\end{figure}

The diagrams we need to compute are shown in Fig.~\ref{bsll03},
where the boxes denote the insertion of the following 
effective non-local Lagrangian
\be
\cL_{\rm eff} = \kappa_F \left[ 
\widetilde{C}_9^{\rm eff} \bar{s} \gamma_\mu  P_L b~\bar{\ell} \gamma^\mu \ell
+\widetilde{C}_{10}^{\rm eff} \bar{s} \gamma_\mu  P_L b~\bar{\ell} \gamma^\mu\gamma_5 \ell 
+\frac{m_b}{e} \widetilde{C}_{7}^{\rm eff} \bar{s} 
\sigma_{\mu\nu} P_R b F^{\mu\nu}\right]~,
\label{eq:effL}
\ee
where 
\be
\kappa_F = \frac{\alpha_{\rm em} G_F V_{ts}^* V_{tb} }{\pi\sqrt{2}}
\ee
and, for simplicity, we have omitted to explicitly show the $s$ dependence 
of $\widetilde{C}_9^{\rm eff}$.
Using  (\ref{eq:effL}) instead of the local Hamiltonian in (\ref{eq:effH}),  
we can easily take into account the (two-loop) IR-divergent 
contributions of four-quark operators, whose one-loop 
$b\to s \ell^+\ell^-$ matrix element is encoded  
in the non-local coefficients 
$\widetilde{C}_i^{\rm eff}$ \cite{Asa1} (see appendix).

\subsection{Virtual corrections}

The amplitude for the three-body process $b(p) \to s(p_s) \ell^+(p_1) \ell^-(p_2)$ 
can be written as 
\bea
\cA_{[3]} &=& \frac{2\kappa_F}{m_b} \left\{ \bar{s} P_R \left[ m_b \gamma^\mu A(s) 
 + (p^\mu + p_s^\nu ) B(s) + O(q^\mu) \right] b \bar{\ell}\gamma_\mu \ell \right. \no \\
&& \qquad  \left.  +  \bar{s} P_R \left[ m_b \gamma^\mu A'(s) 
 + (p^\mu + p_s^\nu ) B'(s) + O(q^\mu) \right] b 
  \bar{\ell}\gamma_\mu \gamma_5  \ell \right\}~, \
\label{eq:A3}
\eea
where $q=p_1+p_2$ and, at the tree level,
\be
A^{(0)} =   \frac{1}{2} \widetilde{C}_9^{\rm eff} + \frac{1}{s} \widetilde{C}_7^{\rm eff}~,\qquad 
B^{(0)} = - \frac{1}{s} \widetilde{C}_7^{\rm eff}~,\qquad  
A'^{(0)} =   \frac{1}{2} \widetilde{C}_{10}^{\rm eff}~,\qquad  
B'^{(0)} =  0~.
\label{eq:A3b}
\ee
Employing the same notation, and working in $d=4-2\eps$ dimensions, 
the renormalized virtual one-loop corrections in Fig.~\ref{bsll03} 
are given by 
\bea
A^{(1)} &=& \frac{ \alpha_s}{3\pi} \Gamma(1+\eps) (1-s)^{-2\eps} 
\left(\frac{m_b^2}{4\pi} \right)^{-\eps} \left\{   
\widetilde{C}_9^{\rm eff} \left[ -\frac{1}{2\eps^2} -\frac{5}{4\eps}
 -\mbox{Li}_2(s)
-\frac{1+2s}{2s} \ln(1-s)-3 \right]\right. \no\\
&& \qquad \left. + \frac{1}{s} \widetilde{C}_7^{\rm eff} \left[ -\frac{1}{\eps^2} -\frac{5}{2\eps} -2 \mbox{Li}_2(s) 
-3\ln(1-s)-10-8\ln\left(\frac{\mu}{m_b}\right)\right] + O(\eps) \right\}~, \no\\
B^{(1)} &=& \frac{ \alpha_s}{3\pi} \Gamma(1+\eps) (1-s)^{-2\eps}
\left(\frac{m_b^2}{4\pi} \right)^{-\eps} \left\{   
\widetilde{C}_9^{\rm eff} \frac{1}{2s} \ln(1-s) \right.  \no \\
&& \qquad \left. + \frac{1}{s} \widetilde{C}_7^{\rm eff} \left[ \frac{1}{\eps^2} +\frac{5}{2\eps} +2 \mbox{Li}_2(s) 
+ \ln(1-s)+10+8\ln\left(\frac{\mu}{m_b}\right)\right] + O(\eps) \right\}~, \no\\
A'^{(1)} &=& \frac{ \alpha_s}{3\pi} \Gamma(1+\eps) (1-s)^{-2\eps}  
\left(\frac{m_b^2}{4\pi} \right)^{-\eps}  
\widetilde{C}_{10}^{\rm eff} \left[ -\frac{1}{2\eps^2} -\frac{5}{4\eps} -\mbox{Li}_2(s)
-\ln(1-s)\right. \no\\ 
&& \qquad \left. -3-\frac{1}{2s}\ln(1-s) + O(\eps) \right]~, \no\\
B'^{(1)} &=& \frac{ \alpha_s}{3\pi} \Gamma(1+\eps) (1-s)^{-2\eps}
\left(\frac{m_b^2}{4\pi} \right)^{-\eps} \widetilde{C}_{10}^{\rm eff} \left[
 \frac{1}{2s} \ln(1-s) \right]~.  \label{eq:A1}
\eea
The above results include also self-energy corrections of the external 
legs, not explicitly shown in Fig.~\ref{bsll03}. 
Effectively, these contributions are included using 
on-shell renormalization conditions. For later comparison, we state  here 
the  corresponding on-shell wave-function renormalization constants for the
external quark fields, where we explicitly separated ultraviolet (UV)
and IR poles:
\bea
\label{fieldmb}
    Z_\psi(m_b) &=& 1 - \frac{\alpha_s}{4\pi} \frac{4}{3} 
\left(\frac{m_b^2}{\mu^2}\right)^{- \epsilon}
        \left( \frac{1}{\epsilon_{\rm UV}} + \frac{2}{\epsilon_{\rm IR}} + 4 \right)~,
\no \\
 Z_\psi(m_s = 0) &=&  1 - \frac{\alpha_s}{4\pi} \frac{4}{3} 
        \left( \frac{1}{\epsilon_{\rm UV}} - \frac{1}{\epsilon_{\rm IR}}\right)~.
\eea
Because of the conservation of vector and axial currents, UV
divergences cancel out completely in the terms proportional to 
$\widetilde{C}_9^{\rm eff}$ and $\widetilde{C}_{10}^{\rm eff}$. 
On the other hand, UV 
divergences proportional to $\widetilde{C}_7^{\rm eff}$ have been eliminated 
by the on-shell mass counterterm  $Z_{m_b}$ and by the 
${\rm \overline{MS}}$ renormalization of the corresponding operator using the  
counterterm $Z_{77}$:
\bea
\label{77mb}
    Z_{77} &=& 1 + \frac{\alpha_s}{4\pi} \frac{16}{3} 
\frac{1}{\epsilon_{\rm UV}}~, \no \\
 Z_{m_b}  &=&  1 - \frac{\alpha_s}{4\pi} \frac{4}{3} 
\left(\frac{m_b^2}{\mu^2}\right)^{- \epsilon}
        \left( \frac{3}{\epsilon_{\rm UV}} +4 \right)~.
\eea
After the UV renormalization
has been performed, we suppress the $\mu$ dependence
corresponding to the IR divergences in order to simplify the notation, 
as can be seen in Eq. (\ref{eq:A1}).

\medskip 

Taking into account the full $d$-dimensional three-body phase space, 
virtual IR corrections to the rate and forward--backward asymmetry 
are obtained by means of Eqs.~(\ref{eq:A3})--(\ref{eq:A1}),
isolating the $O(\alpha_s)$ terms in 
\bea
\label{phasespacevirtual}
{d \Gamma_{[3]}} &=&
\frac{m_b}{2}  \,  \frac{d^{d-1}p_s}{(2\pi)^{d-1} 2E_s} \,
     \frac{d^{d-1}q}{(2\pi)^{d-1} 2q_0}\,  (2\pi)^{d-1} \,
\delta^d(p-p_s-q)  \times \no \\    
&\times&                 \frac{d^{d-1}p_1}{(2\pi)^{d-1}2E_1}\,    \frac{d^{d-1}p_2}{(2\pi)^{d-1}2E_2} \, 
(2\pi)^{d}\, 
\delta^d(q-p_1-p_2)\, d s  \quad \frac{1}{2} \sum_{\rm spins} \frac{1}{3} \sum_{\rm colours} 
\left| \cA_{[3]} \right|^2~
\eea

\subsection{Bremsstrahlung corrections to $R(s)$}
The matrix element of the four-body process $b(p)\to s(p_s) g(k) \ell^+(p_1) \ell^-(p_2)$, 
squared, can in general be decomposed as 
\be
\frac{1}{2} \sum_{\rm spins}  \frac{1}{3} \sum_{\rm colors}
\left| \cA_{[4]} \right|^2 (s,y,x_p,x_s,x_v) 
= H_{\mu\nu}^S L_S^{\mu\nu} + H_{\mu\nu}^A L_A^{\mu\nu}~,
\ee
where  $L_{S,A}^{\mu\nu}$ denote the leptonic tensors
\be
L_S^{\mu\nu} = {\rm tr}\left(\pslash_1 \gamma^\mu \pslash_2 \gamma^\nu \right) \qquad 
{\rm and} \qquad 
L_A^{\mu\nu} = {\rm tr}\left(\pslash_1 \gamma^\mu \pslash_2 \gamma^\nu \gamma_5 \right)~,
\ee
and in addition to $s$ we have introduced the following kinematical variables
\be 
x_p= \frac{k \cdot p}{m^2_b}~, 
\qquad x_s= \frac{k \cdot p_s }{m^2_b}~, \qquad 
y=\frac{p\cdot(p_1-p_2)}{m^2_b}~, \qquad x_v = \frac{k \cdot (p_1-p_2) }{m^2_b}~.
\ee
The asymmetric leptonic tensor $L_A^{\mu\nu}$ vanishes if averaged over the 
full leptonic phase space (symmetric in $p_1 \leftrightarrow  p_2$). 
On the other hand, in the case of $L_S^{\mu\nu}$ we can write 
\bea
I^{\mu\nu}_S &=& (2\pi)^{d} \int \frac{d^{d-1}p_1}{(2\pi)^{d-1}2E_1} \int \frac{d^{d-1}p_2}{(2\pi)^{d-1}2E_2} 
\delta^d(q-p_1-p_2) L_S^{\mu\nu} \no \\
&=& \frac{(1-\eps) (q^2)^{-\eps} }{ \pi^{\frac{1-2\eps}{2}}
2^{3-4\eps} \Gamma\left(\frac{5}{2}-\eps \right)} \left( q^\mu q^\nu - q^2 g^{\mu\nu} \right)~.
\label{leptonic}
\eea
Since $I^{\mu\nu}_S$ depends only on $q$, the variables 
$y$ and $x_v$ do not appear in the calculation of the (symmetric) dilepton spectrum. 

The explicit calculation of the real-emission diagrams in Fig.~\ref{bsll03} leads to 
\bea
&&\!\!\!\!\!\!\!\! \frac{1}{m_b^2} \left( q^\mu q^\nu - q^2 g^{\mu\nu} \right) H_{\mu\nu}^S 
   =  \frac{16\pi}{3}  \kappa_F^2 \alpha_s  \left\{ \frac{}{} (1-s)\left[ -\frac{1}{x_p^2} + \frac{1-s}{x_p x_s} 
 - \frac{2}{x_s}  + (1-\eps) \frac{2x_p}{x_s (1-s)} + \frac{2}{x_p}
 \right] \right.  \no\\
&& \qquad \times \left[ \frac{1 +2s-2\eps s}{2} \left( \left| \widetilde{C}_9^{\rm eff}\right|^2 
+ \left| \widetilde{C}_{10}^{\rm eff} \right|^2 \right) 
+\frac{4+2s-4\eps}{s}  \left| \widetilde{C}_7^{\rm eff} \right|^2 +(6-4\eps) 
\Re\left(\widetilde{C}_7^{\rm eff}\widetilde{C}_9^{\rm eff*} \right)\right] \no\\
&& \quad \qquad + \frac{ x_s (1+2s) -2 x_p }{x_p} \left( \left|\widetilde{C}_9^{\rm eff}\right|^2  +
\left| \widetilde{C}_{10}^{\rm eff} \right|^2\right) 
+ \frac{ 4x_s (s-2+8 x_p )-8x_p(s+4) }{s x_p}  \left| \widetilde{C}_7^{\rm eff} \right|^2 \no \\
&& \left. \quad \qquad - \frac{ 4 x_s +8 x_p }{x_p} \Re\left(\widetilde{C}_7^{\rm eff}\widetilde{C}_9^{\rm eff*} \right)
+O(\eps) \right\}~.
\label{eq:MS41}
\eea
Infrared singularities arise only by terms 
proportional to $1/x_p^2$ or $1/x_s$: for this reason only the $O(\eps)$  
pieces proportional to these terms (between square brackets) have been explicitly shown.
Taking into account the full $d$-dimensional four-body phase space, the  
radiative differential rate can be written as  
\bea
\frac{d \Gamma_{[4]}}{ds} &=&
\frac{m_b}{2} (2\pi)^{d-1} \int \frac{d^{d-1}p_s}{(2\pi)^{d-1} 2E_s} \int \frac{d^{d-1}k}{(2\pi)^{d-1} 2E_g} 
\int \frac{d^{d-1}q}{(2\pi)^{d-1} 2q_0} 
\delta^d(p-p_s-k-q) I_S^{\mu\nu} H^S_{\mu\nu} \no \\ 
&=& \frac{ C_S }{16\pi^2} s^{-\eps} (1-s)^{3-4\eps} \int_0^1 dx  (1-x)^{1-2\eps} x^{1-2\eps}  
    \int_0^1 d \omega (1-\omega)^{-\eps} \omega^{-\eps}  \no\\
&&  \times [1-x\omega(1-s)]^{-2+2\eps} 
    \left[    \frac{1}{m_b^2} \left( q^\mu q^\nu - q^2 g^{\mu\nu} \right) H_{\mu\nu}^S    \right]~,
 \qquad\quad  \label{eq:full_G} 
\eea
where 
\be
C_S = \frac{(1-\eps) m_b^{5-6\eps} }{\pi^{2-3\eps}2^{9-10\eps}
\Gamma\left(1-\eps\right)
\Gamma\left(\frac{5}{2}-\eps\right)\Gamma\left(\frac{3}{2}-\eps\right)}~.
\ee
The integration variables in Eq.~(\ref{eq:full_G}) are defined by 
\be
\omega = \frac{1}{2} (1-\cos\theta_{sk})~, \qquad \qquad 
x=\frac{ 2 p\cdot k}{ m_b^2-q^2}~,
\label{eq:intvar}
\ee
where $\theta_{sk}$ is the angle between gluon and strange-quark momenta
in the $B$ rest frame, so that 
\be
x_p(s,x) = \frac{x(1-s)}{2}~, \qquad 
x_s(s,x,\omega) = \frac{ x  \omega (1-x) (1-s)^2 }{2 [1-x\omega(1-s)]}~.
\label{eq:xsp_expl}
\ee
Performing explicitly the integrals in $x$ and $\omega$ we 
obtain
\bea
&&\!\!\! \frac{d \Gamma_{[4]}}{ds} = \frac{\alpha_s}{3\pi}  \kappa_F^2  C_S 
s^{-\eps} (1-s)^{2-4\eps}  \no\\
&&\ \times \left\{ \left[ \frac{2}{\eps^2} +\frac{5}{\eps} -4\mbox{Li}_2(s) 
  -4\ln(s)\ln(1-s)-\pi^2  +\frac{35}{2} 
+\frac{1}{1-s} -\frac{s(2-3s)}{(1-s)^2}\ln(s) \right]    \right. \no\\
&&\ \ \ \times \left[ \frac{1 +2s-2\eps s}{2} 
\left( \left| \widetilde{C}_9^{\rm eff}\right|^2 
+ \left| \widetilde{C}_{10}^{\rm eff} \right|^2 \right) 
+\frac{4+2s-4\eps}{s}  \left| \widetilde{C}_7^{\rm eff} \right|^2 +(6-4\eps) 
\Re\left(\widetilde{C}_7^{\rm eff}\widetilde{C}_9^{\rm eff*} \right)\right] \qquad \no\\
&& \qquad +\frac{s(4s^2+10s-4)\ln(s)-10s^3+7s^2+6s-3}{4(1-s)^2}
\left( \left|\widetilde{C}_9^{\rm eff}\right|^2 + \left| \widetilde{C}_{10}^{\rm eff} \right|^2\right)  \no\\
&& \qquad +\frac{s(6s^2+24s-24)\ln(s)-19s^3+18s^2+15s-14}{3s(1-s)^2}\left| \widetilde{C}_{7}^{\rm eff} \right|^2 \no\\
&& \qquad \left. 
-\frac{s(2s+12)\ln(s)-9s^2+4s+5}{(1-s)^2}   \Re\left(\widetilde{C}_7^{\rm eff}\widetilde{C}_9^{\rm eff*}\right) \right\}~.
\label{eq:R_BR} 
\eea

In order to construct an IR-safe observable,
$d\Gamma_{[4]}/ds$ has to be combined with the $\cO(\alpha_s)$ 
virtual corrections in the non-radiative process.
The differential decay rate for the latter can be written as 
\bea
\frac{d \Gamma_{[3]}}{ ds} &=& \kappa_F^2 C_S 
s^{-\eps} (1-s)^{2-2\eps}  \Gamma(1-\eps) \left( \frac{m_b}{ 4\pi} \right)^\eps \no \\
& \times& \!\!\!\!\! 
\left\{ 2 A^2 [1+ 2 s (1-\eps)] + 4AB(1-s) + 2B^2(1-s)^2 + (A,B \leftrightarrow A',B') \right\} \qquad 
\eea
in terms of the reduced amplitudes ($A,~B$) and ($A',~B'$) defined in (\ref{eq:A3}).
Using Eqs.~(\ref{eq:A3b})--(\ref{eq:A1}) and isolating the $\cO(\alpha_s)$
terms we then obtain 
\bea
&&\!\!\! \frac{d \Gamma^{(1)}_{[3]}}{ds} = \frac{\alpha_s}{3\pi} \kappa_F^2 C_S 
s^{-\eps} (1-s)^{2-4\eps}  \Gamma(1+\eps)\Gamma(1-\eps) \no\\
&&\ \times \left\{ -\frac{1 +2s-2\eps s}{2}\left( \left| \widetilde{C}_9^{\rm eff}\right|^2 
+ \left| \widetilde{C}_{10}^{\rm eff} \right|^2 \right) 
\left[\frac{2}{\eps^2} +\frac{5}{\eps} +4\mbox{Li}_2(s)
+\frac{2+4s}{s} \ln(1-s)+12 \right] \right. \no\\
&&\quad\  - \frac{4+2s-4\eps}{s}  \left| \widetilde{C}_7^{\rm eff} \right|^2 
\left[\frac{2}{\eps^2} +\frac{5}{\eps} +4 \mbox{Li}_2(s) 
+4\ln(1-s)+20+16\ln\left(\frac{\mu}{m_b}\right) \right] \no\\
&&\quad\  -(6-4\eps) \Re\left(\widetilde{C}_7^{\rm eff}\widetilde{C}_9^{\rm eff*} \right)
\left[\frac{2}{\eps^2} +\frac{5}{\eps} +4 \mbox{Li}_2(s) 
+\frac{1+4s}{s}\ln(1-s)+16+8\ln\left(\frac{\mu}{m_b}\right) \right] \no\\
&&\left. \ +\frac{\ln(1-s)}{s} \left[ (1-s) \left( \left| 
 \widetilde{C}_9^{\rm eff}\right|^2 + 
 \left| \widetilde{C}_{10}^{\rm eff} \right|^2 \right)  -4(4-s)\left| 
\widetilde{C}_7^{\rm eff} \right|^2  
 -2(1+2s) \Re\left(\widetilde{C}_7^{\rm eff}\widetilde{C}_9^{\rm eff*} \right)\right] 
\right\}
\no\\
\label{eq:R_vir} 
\eea
It is straightforward to check that all divergences cancel in the sum 
of Eq.~(\ref{eq:R_BR}) and Eq.~(\ref{eq:R_vir}). Combining these two equations 
we can finally obtain explicit expression for the $\tau_i(s)$ 
and $\sigma_i(s)$ functions defined in Eq.~(\ref{NNLLDIMS}). 

As mentioned before, we define the universal functions 
$\sigma_{7,9}(s)$ such that the non-universal terms 
$\tau_{77}(s)$ and $\tau_{99}(s)$ vanish for $s\to 1$. 
This is because in the $s\to 1$ limit
only the soft component of the radiation survives and, according 
to Low's theorem, the latter gives rise to a correction proportional 
to the tree-level matrix. We then define 
\bea
  \sigma_9(s) &=& \sigma(s) + \frac{3}{2}~, \qquad\qquad  
  \sigma_7(s) \ = \ \sigma(s) + \frac{1}{6} - \frac{8}{3}\ln\left( \frac{\mu}{m_b} \right)~, 
\no\\
  \sigma(s) &=& - \frac{4}{3} \mbox{Li}_2(s) - \frac{2}{3} \ln(s) \ln(1-s)
  -\frac{2}{9}\pi^2 -\ln(1-s)-\frac{2}{9}(1-s)\ln(1-s)~. \qquad 
\label{eq:sigmai}
\eea
With this choice the $\tau_i(s)$ are given by 
\bea
\tau_{77}(s) &=& - \frac{2}{9(2+s)} \left[   2(1-s)^2\ln(1-s)
+\frac{6s(2-2s-s^2)}{(1-s)^2}\ln(s) +\frac{11-7s-10s^2}{(1-s)} \right]~, \no\\
 \tau_{99}(s) &=& 
-\frac{4}{9(1+2s)} \left[ 2(1-s)^2\ln(1-s) 
+\frac{3s(1+s)(1-2s)}{(1-s)^2}\ln(s)+\frac{3(1-3s^2)}{1-s} \right]~, \no \\
\tau_{79}(s) &=&  - \frac{4(1-s)^2}{9s} \ln(1-s)
-\frac{4s(3-2s)}{9(1-s)^2}\ln(s) -\frac{2(5-3s)}{9(1-s)}~,
\label{eq:tau_R}
\eea
and all vanish at $s=1$.

These results are in complete agreement with those recently 
obtained by Asatrian et al.~\cite{Asa1}. Indeed 
the $\omega_i(s)$ of Ref.~\cite{Asa1} can be written as
$\omega_7 = \sigma_7 + \tau_{77}/2$,
$\omega_9 = \sigma_9 + \tau_{99}/2$,
$\omega_{79} = (\sigma_7 + \sigma_9 + \tau_{79})/2$.

\subsection{Virtual and bremsstrahlung corrections to $A_{\rm FB}(s)$}

A use of na\"\i ve dimensional regularization is not allowed
in the case of the forward--backward asymmetry
because of the ambiguities arising for $d\not=4$
in the definition of $\gamma_5$ (or in the trace of $L_{A}^{\mu\nu}$);
in the case of the decay rate, the problematic $\gamma_5$ 
contribution vanishes because of the $p_1 \leftrightarrow p_2$ permutation 
symmetry of the leptonic phase space [see Eq. (\ref{leptonic})]. 
To circumvent this problem, we employ the following regularization
scheme: we treat the Dirac algebra of IR-divergent pieces 
strictly in four dimensions (dimensional reduction). 
This of course simplifies 
the calculation of the real emission, where only IR 
divergences appear, but it slightly complicates the virtual 
corrections, where UV divergences also occur. The latter, 
which do not involve any $\gamma_5$ ambiguity, are 
still computed in na\"\i ve dimensional regularization
and renormalized as discussed at the beginning of this section.
We stress that, at this level of the perturbative expansion, 
this hybrid regularization scheme is gauge invariant.

Within the virtual corrections we therefore need to 
strictly separate IR and UV divergences. As a result,
the on-shell wave-function-renormalization factors 
given in  Eqs. (\ref{fieldmb}) within na\"\i ve dimensional 
regularization are partially changed. While the IR part of 
$Z_\psi(m_b)$ is not modified,  $Z_\psi(m_s = 0)$ is different
within our hybrid dimensional scheme:
\bea
\label{fieldmbhybrid}
    Z_\psi(m_b) &=& 1 - \frac{\alpha_s}{4\pi} \frac{4}{3} 
\left(\frac{\mu}{m}\right)^{2\epsilon}
        \left( \frac{1}{\epsilon_{\rm UV}} + \frac{2}{\epsilon_{\rm IR}} + 4 \right)~,
\no \\
 Z_\psi(m_s = 0) &=&  1 - \frac{\alpha_s}{4\pi} \frac{4}{3} 
        \left( \frac{1}{\epsilon_{\rm UV}} - \frac{1}{\epsilon_{\rm IR}}  - 1 \right)~.
\eea
Being of pure UV nature, the other $Z$ factors remain unchanged.
On the amplitude level, the change of $Z_\psi(m_s = 0)$ turns out to be 
the only change of virtual corrections. Taking this effect into account, 
Eqs.~(\ref{eq:A1}) are modified as follows:
\bea
A^{(1)}_{D4} &=& A^{(1)} +
\frac{ \alpha_s}{3\pi} \Gamma(1+\eps) (1-s)^{-2\eps} 
\left(\frac{m_b^2}{4\pi} \right)^{-\eps} \left[
\frac{1}{4}\widetilde{C}_9^{\rm eff} 
+ \frac{1}{2s}\widetilde{C}_7^{\rm eff}  \right]~, \no\\
B^{(1)}_{D4} &=& B^{(1)} +
\frac{ \alpha_s}{3\pi} \Gamma(1+\eps) (1-s)^{-2\eps} 
\left(\frac{m_b^2}{4\pi} \right)^{-\eps} \left[ - \frac{1}{2s}\widetilde{C}_7^{\rm eff} 
\right]~,  \no\\
A'^{(1)}_{D4} &=& A'^{(1)} +
\frac{ \alpha_s}{3\pi} \Gamma(1+\eps) (1-s)^{-2\eps} 
\left(\frac{m_b^2}{4\pi} \right)^{-\eps} \left[   
\frac{1}{4}\widetilde{C}_9^{\rm eff}  \right]~,  \qquad 
B'^{(1)}_{D4} = B'^{(1)}~. \qquad 
\label{eq:A1D}
\eea
As a consistency check, we have explicitly verified 
that the finite corrections to $R(s)$,
computed in the hybrid scheme, are exactly the same as in 
Eqs.~(\ref{eq:sigmai})--(\ref{eq:tau_R}).
To this purpose, the bremsstrahlung  term with four-dimensional Dirac algebra 
is obtained  by neglecting all the $O(\eps)$ terms
in Eq.~(\ref{eq:MS41}).

\medskip

\begin{figure}[t]
\begin{center}
\epsfig{file=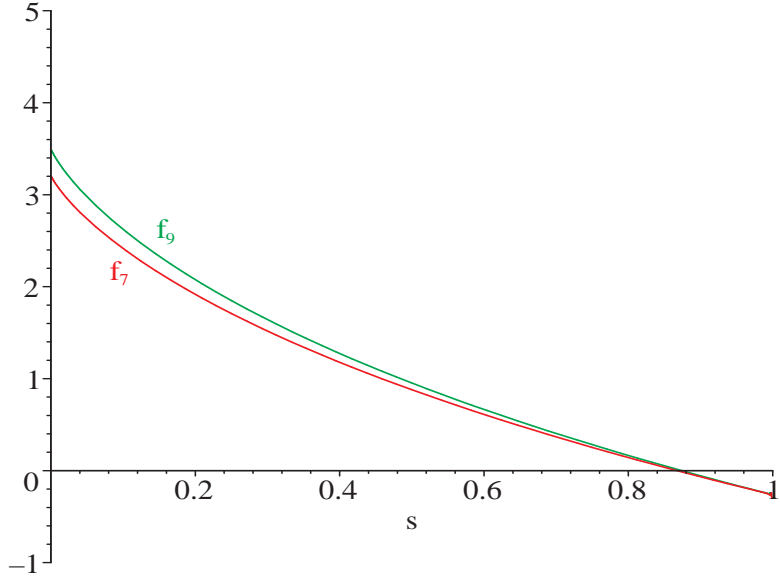,width=16cm,height=9cm}
\end{center}
\caption{The functions $f_7(s)$ and $f_9(s)$.}
\label{fig:f79}
\end{figure}

The asymmetric part of 
the squared matrix element of the four-body process 
--- relevant to $A_{\rm FB}(s)$ --- 
computed with four-dimensional Dirac 
algebra, is given by
\bea
\frac{1}{m_b^2}  H_{\mu\nu}^A L_A^{\mu\nu}~
 &=&  \frac{128\pi}{3}  \kappa_F^2 \alpha_s 
\left\{ y \left[ -\frac{1}{x_p^2} + \frac{1-s}{x_p x_s} 
 - \frac{2}{x_s} + \frac{2x_p}{x_s(1-s)} \right]
\Re\left[ \widetilde{C}_{10}^{\rm eff*} \left( 
s \widetilde{C}_9^{\rm eff}
+2 \widetilde{C}_7^{\rm eff} \right)\right] \right.  \no\\
&&  +  s \left[ \frac{2yx_p-x_v(1-s)}{2 x_p x_s(1-s)}(1-s-2x_p)
+ \frac{y}{x_p} +\frac{x_v}{x_p^2}  \right]
\Re\left( \widetilde{C}_{10}^{\rm eff*} \widetilde{C}_9^{\rm eff} \right) 
 \no\\
&&  + \frac{1}{s}  \left[ \frac{2yx_p-x_v(1-s)}{2 x_p x_s(1-s)}
\left(2s(1-s-2x_p)+x_p(1-s)\right)
+ \frac{y}{x_p}(x_s+2s-2x_p)  \right. \no\\
&& \left.\left. \quad +\frac{x_v}{2 x_p^2} \left( 4s-2x_sx_p+(1-3s)x_p+2x_p^2
\right) \right]  
\Re\left( \widetilde{C}_{10}^{\rm eff*} \widetilde{C}_7^{\rm eff} \right) 
\right\}~.
\label{eq:MA41}
\eea
As can be noted, $H_{\mu\nu}^A L_A^{\mu\nu}$
 is linear in the leptonic asymmetric 
variables $y$ and $x_v$. The latter can be written as 
\bea
y &=& - \cos\theta_\ell \left[ \left(\frac{p\cdot q}{m_b^2}\right)^2 -
\frac{q^2}{m_b^2} \right]^{1/2}, \no\\
x_v &=& y~\frac{p\cdot k}{m_b^2} \left[ 1+  (p\cdot q)
\frac{ (p\cdot p_s)\cos\theta_{sk}  +  (p\cdot k) }{ (p\cdot q)^2 -  q^2 m_b^2 }
 +O(\sin\phi_\ell) \right]~, \qquad 
\label{eq:xv_wrong1}
\eea
where $\theta_\ell$ is the angle between $\ell^+$ and $B$ 
momenta in the dilepton centre-of-mass frame and $\phi_\ell$
the corresponding azimuthal angle. The integrals of $y$ and 
$x_v$ over the leptonic phase space, weighted by
$\mbox{sgn}(\cos\theta_\ell)$, can be expressed in terms of $s$
and the two integration variables $x$ and $\omega$ defined in
Eq.~(\ref{eq:intvar}):
\bea
\int \frac{d^{d-1}p_1}{2E_1}\!\!\!\!\!\!\!\!  &&\!\!\!\! \int \frac{d^{d-1}p_2}{2E_2} 
  \frac{\delta^d(q-p_1-p_2)}{(2\pi)^{d-2}} \mbox{sgn}(\cos\theta_\ell)
 \left\{ \ba{c} 
y \\ x_v \ea \right\} = \frac{ (q^2)^{-\eps} (1-\eps)^{-1}  }{ \pi^{1-\eps}
2^{4-4\eps} \Gamma\left(1-\eps \right)}  \left\{ \ba{c} 
{\overline y}(s, x,\omega) \\ {\overline x_v}(s, x,\omega) \ea \right\} \no\\
{\overline y}(s, x,\omega)\!\! &=&\!\! - \frac{(1-s)\beta(s, x,\omega) }{2[ 1 -x(1-s)\omega]}~, \no\\
{\overline x_v }(s, x,\omega)\!\! &=&\!\! 
 {\overline y} x_p \left[ 1 + (1+s-2x_s) \frac{
 (1-s+2x_s)(1-2\omega)+4\omega x_p}{(1-s)^2-4(1+s)x_s+4x_s^2} 
 \right]~,
\label{eq:xv_wrong2}
\eea
where $x_{s,p}$ are given as in Eq.~(\ref{eq:xsp_expl}) and 
\be
\beta(s, x,\omega) =  \sqrt{ 1 -x\omega\left[ 4(1-x)+2x(1-s)(1-2\omega)
+x^2\omega (1-s) (4-x(1-s)) \right] }~.
\ee 
Taking into account also the phase-space integration over 
the hadronic variables, the regularized 
bremsstrahlung contribution to the FB asymmetry can be written as  
\bea
\frac{d \Gamma^A_{[4]}}{ds} &=&  \int d\cos\theta_\ell ~
 \frac{d^2 \Gamma_{[4]} }{d s ~ d\cos\theta_\ell}
\mbox{sgn}(\cos\theta_\ell) = \frac{ C_A }{16\pi^2} s^{-\eps} (1-s)^{3-4\eps}  \int_0^1 dx 
(1-x)^{1-2\eps} x^{1-2\eps} \no \\
&&   \int_0^1 d \omega (1-\omega)^{-\eps} \omega^{-\eps}[1-x\omega(1-s)]^{-2+2\eps}
\left[\frac{1}{m_b^2}  H_{\mu\nu}^A L_A^{\mu\nu}\right]_{ 
y \to {\overline y}(s, x,\omega), x_v \to {\overline x_v}(s, x,\omega) }
\quad\qquad  \label{eq:full_A} 
\eea
where 
\be
C_A = \frac{ m_b^{5-6\eps} }{ \pi^{\frac{5}{2}-3\eps}2^{10-10\eps} (1-\eps)
\Gamma\left(1-\eps\right)\Gamma\left(1-\eps\right) \Gamma\left(\frac{3}{2}-\eps\right)}~.
\ee
Because of the square root in $\beta(s, x,\omega)$, 
the integrals on $x$ and $\omega$ appearing in (\ref{eq:full_A})
are rather involved and we have not been able to perform them analytically.
However, divergent contributions arise only  in the limit 
$\overline{y} \to - (1-s)/2$ from the terms
in the first line of (\ref{eq:MA41}). 
Computing analytically only the latter, we can write 
\bea
&&\!\!\! \frac{d \Gamma^A_{[4]}}{ds} = - \frac{4\alpha_s}{3\pi}  \kappa_F^2  C_A 
s^{-\eps} (1-s)^{2-4\eps}  \no\\
&&\ \times \left\{ \left[ \frac{2}{\eps^2} +\frac{5}{\eps} -4\mbox{Li}_2(s) 
  -4\ln(s)\ln(1-s)-\pi^2  +\frac{25}{2} 
+\frac{1}{1-s} -\frac{s(6-7s)}{(1-s)^2}\ln(s) \right]    \right. \no\\
&&\qquad \times  \Re\left[ \widetilde{C}_{10}^{\rm eff*} \left( 
s \widetilde{C}_9^{\rm eff} 
+2 \widetilde{C}_7^{\rm eff} \right)\right] \left. 
 + f_{9}(s)  \Re\left( s \widetilde{C}_{10}^{\rm eff*} \widetilde{C}_9^{\rm eff} \right)
 + f_{7}(s)  \Re\left( 2 \widetilde{C}_{10}^{\rm eff*} \widetilde{C}_7^{\rm eff} \right) 
 \frac{}{}  \right\}~, \qquad\quad
\label{eq:A_BR} 
\eea 
where $f_{7,9}(s)$ denotes regular functions (for $s\not=0$), which we evaluate by means 
of numerical methods. The numerical results obtained for  $f_{7,9}(s)$ are shown in 
Fig.~\ref{fig:f79}. 

\medskip

The cancellation of IR divergences is obtained 
by combining $d \Gamma^A_{[4]}/ds$ with the $\cO(\alpha_s)$
terms in  
\be
\frac{d \Gamma^A_{[3]}}{ ds} = - \kappa_F^2 C_A 
s^{1-\eps} (1-s)^{2-2\eps}  \Gamma(1-\eps) \left( \frac{m_b}{ 4\pi} \right)^\eps \left[16 A_{D4} A_{D4}' \right]~.
\ee
Notice that, as explicitly indicated, in this case one should use the one-loop 
expressions of $A$ and $A'$ computed in the hybrid scheme, given in 
Eqs.~(\ref{eq:A1D}). Isolating the $\cO(\alpha_s)$
terms we get
\bea
&&\!\!\! \frac{d \Gamma^{A(1)}_{[3]}}{ds} = \frac{4\alpha_s}{3\pi}  \kappa_F^2  C_A 
s^{-\eps} (1-s)^{2-4\eps} \Gamma(1+\eps)\Gamma(1-\eps)  \no\\
&&\ \times \left\{ \left[ \frac{2}{\eps^2} +\frac{5}{\eps} +4\mbox{Li}_2(s) 
  +\frac{2+4s}{s}\ln(1-s)+11   \right] 
  \Re\left( s \widetilde{C}_{10}^{\rm eff*} \widetilde{C}_9^{\rm eff} \right) \right.  \no\\
&& \quad \left.  + \left[ \frac{2}{\eps^2} +\frac{5}{\eps} +4\mbox{Li}_2(s) 
  +\frac{1+5s}{s}\ln(1-s)+15 +8\ln\left(\frac{\mu}{m_b}\right)   \right] 
  \Re\left( 2 \widetilde{C}_{10}^{\rm eff*} \widetilde{C}_7^{\rm eff} \right) 
 \frac{}{}  \right\} \qquad 
\label{eq:A_Vir} 
\eea 
and by means of Eqs.~(\ref{eq:A_BR}) and (\ref{eq:A_Vir}) we finally obtain
\bea
\tau_{910}(s) &=& - \frac{5}{2}+\frac{1}{3(1-s)}-\frac{s(6-7s)}{3(1-s)^2}\ln(s) 
 -\frac{2}{9s}(3-5s+2s^2)\ln(1-s) +\frac{1}{3}f_9(s)~, \qquad \no \\
\tau_{710}(s) &=& - \frac{5}{2}+\frac{1}{3(1-s)}-\frac{s(6-7s)}{3(1-s)^2}\ln(s) 
 -\frac{1}{9s}(3-7s+4s^2)\ln(1-s) +\frac{1}{3}f_7(s)~. \qquad 
\label{eq:tau_FB}
\eea

\section{Phenomenological analysis}

\begin{figure}[t]
\begin{center}
\epsfig{file=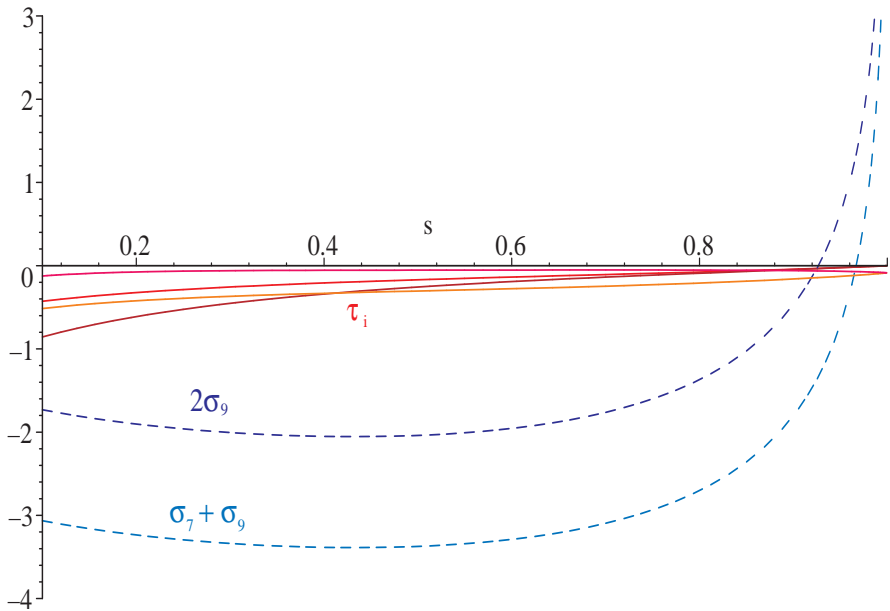,width=15cm}
\end{center}
\caption{The functions $\sigma_i(s)$ (dashed lines) and $\tau_i(s)$
(full lines). At $s=0.1$ the $\tau_i(s)$ functions are 
ordered as follows (from top to bottom): $\tau_{910}$,  $\tau_{79}$,
$\tau_{710}$, and $\tau_{99}$ (almost identical to $\tau_{77}$, 
not explicitely shown). }
\label{fig:sandt}
\end{figure}

The results for $\sigma_i(s)$ and $\tau_i(s)$ functions,
which encode bremsstrahlung and virtual IR corrections 
to $R(s)$ and $A_{\rm FB}(s)$, and which we have determined 
in the previous section, are summarized in Fig.~\ref{fig:sandt}.
As anticipated, in all cases, except for very small values of $s$,
the universal contribution of the $\sigma_i(s)$ is 
largely dominant with respect to the non-universal 
contribution of the $\tau_i$ terms.\footnote{~To simplify the comparison,
in Fig.~\ref{fig:sandt} we have plotted the  $\tau_i(s)$
functions and the two combinations  
$2\sigma_9(s)$ and $\sigma_9(s)+\sigma_7(s)$
which appear in the physical observables.} The 
natural scale of these corrections is 
$\sigma_i \alpha(m_b)/\pi = O(10\%)$, and they thus represent 
a numerically important effect both in the rate and in the FB asymmetry. 

A detailed phenomenological analysis of $R(s)$ and $A_{\rm FB}(s)$,
including all NNLL effects, will be presented elsewhere,
together with an independent calculation of the two-loop 
matrix-element corrections \cite{in_prog}, which extends existing 
calculations also to the high-dilepton-mass region.  
Here we shall limit ourselves to 
analysing how the zero of the FB asymmetry ($s_0$) is modified 
by the inclusion of bremsstrahlung and virtual IR corrections.
As is well known, this quantity, defined by $A_{\rm FB}(s_0)=0$, 
is particularly interesting to determine relative sign and 
magnitude of the Wilson coefficients $C_7$ and $C_9$.  

Employing the counting rule of Ref.~\cite{Asa1}, i.e. treating the  
formally $O(1/\alpha_s)$ term of $\widetilde C_9^{\rm eff}$
as $O(1)$ (see discussion at the end of Section 3), 
the lowest-order value of $s_0$ (formally derived by the NLL 
expression of $A_{\rm FB}$) is determined by the solution of 
\be
s_0 \widetilde C_9^{\rm eff}(s_0) + 2 \widetilde C_7^{\rm eff} = 0~.
\ee
Using the values of the effective coefficients at $\mu=5$ GeV 
reported in Ref.~\cite{Asa1}, this leads to 
\be
s^{\rm NLL}_0 = 0.14 \pm 0.02~,
\label{eq:s0LO}
\ee
where the error is determined by the 
scale dependence ($2.5~\rm{ GeV} \leq \mu \leq 10$~GeV). 

Neglecting the small contribution of $\delta_{\rm FB}$
[see Eq.~(\ref{NNLLAFB})],
the next-to-leading order equation for $s_0$ reads 
\be
s_0 \Cnew_9(s_0)\left(1 +  \frac{\alpha_s}{\pi} \tau_{910}(s_0)\right)
+ 2 \Cnew_7(s_0)\left(1 +  \frac{\alpha_s}{\pi} \tau_{710}(s_0)\right) = 0~,
\ee
which leads to 
\be
s^{\rm NNLL}_0 = 0.162 \pm 0.008~.
\label{eq:s0NLO}
\ee
In this case the variation of the result induced by the scale dependence is 
accidentally very small (about $\pm 1\%$ for $2.5~\rm{ GeV} \leq \mu \leq 10$~GeV)
and we believe it cannot be used as a good estimate of higher-order corrections: 
taking into account the separate scale variation of both $\Cnew_9$ and  $\Cnew_7$,
and the charm-mass dependence, we estimate a conservative overall error
on $s_0$ of about $5 \%$.

\begin{figure}[t]
\begin{center}
\epsfig{file=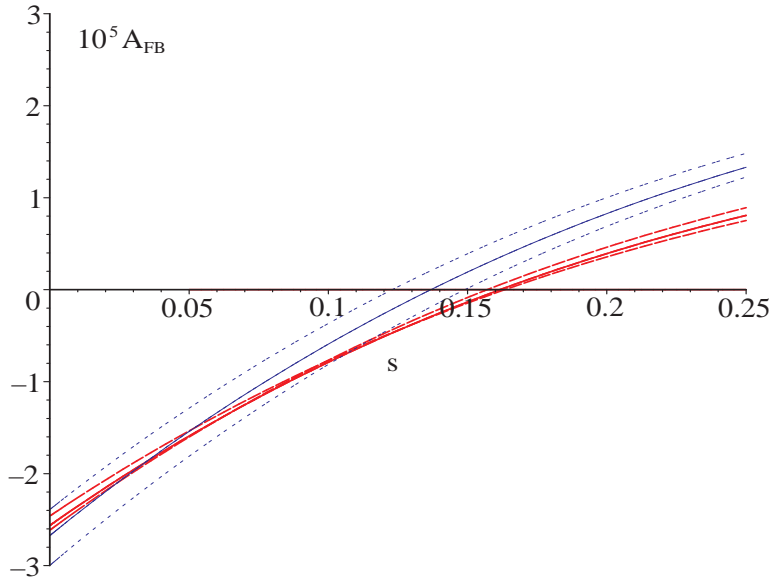,width=16cm,height=9cm}
\end{center}
\caption{Comparison between NNLL and  NLL results for 
$A_{\rm FB}(s)$ in the low $s$ region. 
The three thick (red) lines are the NNLL  
predictions for $\mu=5$~GeV (full),
and $\mu=2.5$ and 10 GeV (dashed); the dotted (blue) curves 
are the corresponding NLL results. All curves have been 
obtained for $m_c/m_b=0.29$.}
\label{fig:AFB}
\end{figure}

We recall that the {\it new} effective Wilson coefficients 
defined in (\ref{effmod}) include not only purely UV virtual terms, 
but also universal bremsstrahlung and corresponding IR virtual 
corrections (parametrized in $\sigma_i$). The global effect of 
the bremsstrahlung and virtual IR corrections we have evaluated 
is that of reducing by $\sim 11\%$ the large ($\sim 27\%$) shift between 
$s^{\rm NNLL}_0$ and $s^{\rm NLL}_0$ induced by the NNLL UV virtual 
terms:\footnote{A similar increase of the position of the zero in the 
FB asymmetry at the $\sim 30\%$ level has been 
found in the exclusive $B \rightarrow V$ channel \cite{martin}. In the inclusive
channel the effect turns out to be somehow reduced by the inclusion
of the bremsstrahlung corrections, which are of course absent in the
exclusive mode.} 
\be
\left. \frac{\delta s_0}{s_0} \right|_{\sigma_i, \tau_i} = \frac{\alpha_s}{\pi} \left[
\sigma_7(s_0) - \sigma_9(s_0) + \tau_{710}(s_0) - \tau_{910}(s_0) \right]
= -0.11~.
\ee
On the other hand, it is clear that the reduction of the error 
(or the reduction of the scale dependence) in Eq.~(\ref{eq:s0NLO}) 
is only due to the inclusion of the UV virtual corrections 
computed in Ref.~\cite{Asa1}. An illustration of the shift of the 
central value and the reduced scale dependence 
between NNL and NNLL expressions of $A_{\rm FB}(s)$, in the low $s$
region, is presented in Fig.~\ref{fig:AFB}. 
We note that in this region the nonperturbative  
$1/m_b^2$ and $1/m_c^2$ corrections to $A_{\rm FB}(s)$ 
are very small \cite{Alineu,Rey,BI} and can safely neglected 
compared to the uncertainties of the partonic calculation.

Interestingly, at this level of precision, the result obtained 
using the ordinary LL expansion is very similar to the 
one obtained using the phenomenological counting rule of Ref.~\cite{Asa1}.
Indeed, at a pure NNLL level, i.e.~neglecting all $O(\alpha_s)$ 
terms in $\Cnew_7$ and retaining the $O(\alpha^{-1}_s \times \alpha_s)$ 
terms of  $\Cnew_9$, we also find as central value 
$s^{\rm NNLL}_0 = 0.16$.

\section{Summary}

The inclusive $B \rightarrow X_s \ell^+\ell^-$ transition,
which is starting to be accessible at $B$ factories \cite{Bsll_exp}, 
represents a new source of theoretically clean observables, 
complementary to the $B \rightarrow X_s \gamma$ rate.
In particular, kinematic observables such as the invariant-dilepton-mass 
spectrum and the lepton forward--backward asymmetry
in \mbox{$B \rightarrow X_s \ell^+\ell^-$}, provide a clean information 
on short-distance couplings not accessible in $B \rightarrow X_s \gamma$.

In the present paper we calculated bremsstrahlung and 
corresponding virtual corrections
in  $B \rightarrow X_s \ell^+\ell^-$.
We used a full dimensional regularization scheme of 
infrared divergences (both soft and collinear ones), which 
also avoids any ambiguity related to the definition of $\gamma_5$. 

For the dilepton-invariant-mass spectrum, these contributions 
have already been evaluated in Refs.~\cite{Asa1,Asa2}, using a 
different technique. Our results are in complete 
agreement with those. We also presented the first computation 
of the soft-gluon (and corresponding virtual) corrections 
in the FB asymmetry, with the help of which we evaluated 
this observable systematically to NNLL precision for the first time. 

The new contributions are rather important and significantly 
improve the sensitivity of the inclusive $B \rightarrow X_s l^+ l^-$ 
decay in testing extensions of the Standard Model in the sector of flavour 
dynamics. In particular, the corrections we have computed shift by about 
10\% the position of the zero of the FB asymmetry. The complete effect of NNLL
contributions to this interesting observable adds up to a $16\%$ shift compared 
with the  NLL result, with a residual error due to higher-order terms  
reduced at the 5\% level.

\section*{Acknowledgments} 
We thank Gerhard Buchalla for many interesting discussions at an 
early stage of this work. 

\section*{Added Note}
After this work was completed, an independent 
calculation of NNLL corrections to the FB asymmetry appeared 
\cite{CH_new}. We find complete agreement with the results 
of Ref.~\cite{CH_new} and we thank Christoph Greub for useful 
correspondence, which allow us to correct a 
typo in Eqs.~(\ref{eq:xv_wrong1}) (which affected 
the $s\to 0$ behaviour of the plot in Fig.~\ref{fig:f79}).

\setcounter{figure}{0}
\setcounter{equation}{0}
\renewcommand{\theequation}{A.\arabic{equation}}
\renewcommand{\thefigure}{A\arabic{figure}}
\section*{Appendix: Auxiliary definitions}

$\bullet \,$  The function $h(z)$ describing next-to-leading order QCD corrections
to the semileptonic decay [see Eq.~(\ref{semi})] is given by \cite{NirNir}:
\bea
\label{nirfunction}
    h(z) = & -(1-z^2) \, \left( \frac{25}{4} - \frac{239}{3} \, z + \frac{25}{4} \, z^2 \right) + z \, \ln(z)
        \left( 20 + 90 \, z -\frac{4}{3} \, z^2 + \frac{17}{3} \, z^3 \right)
        \nonumber \\ &
        + z^2 \, \ln^2(z) \, (36+z^2) + (1-z^2) \, \left(\frac{17}{3} - \frac{64}{3} \, z +
        \frac{17}{3} \, z^2 \right) \, \ln (1-z)
        \nonumber \\ &
        - 4 \, (1+30 \, z^2 + z^4) \, \ln(z) \ln(1-z) -(1+16 \, z^2 +z^4)  \left( 6 \, \mbox{Li}(z) - \pi^2 \right)
        \nonumber \\ &
        - 32 \, z^{3/2} (1+z) \left[\pi^2 - 4 \, \mbox{Li}(\sqrt{z})+ 4 \, \mbox{Li}(-\sqrt{z})
        - 2 \ln(z) \, \ln \left( \frac{1-\sqrt{z}}{1+\sqrt{z}} \right) \right]~.
\nonumber
\eea

\begin{figure}
\begin{center}
\epsfig{file=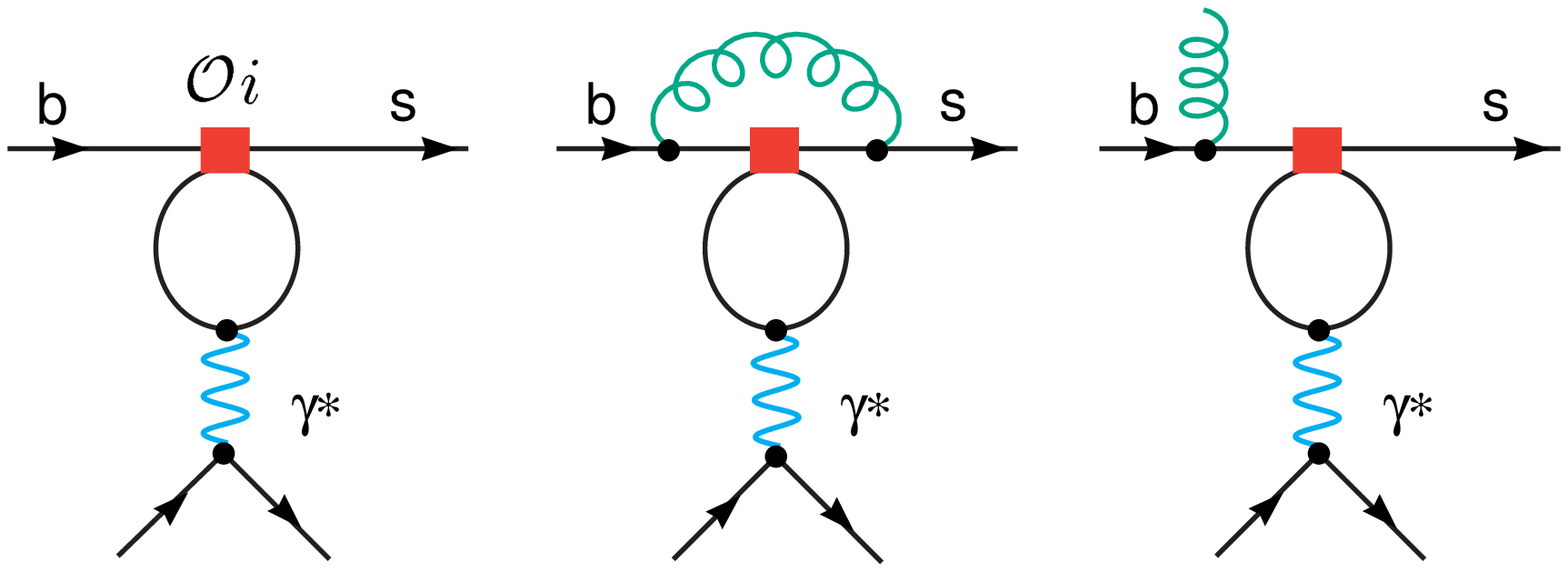,width=10cm}
\end{center}
\caption{Examples of virtual and bremsstrahlung 
contributions of the four-quark operators ${\cal O}_{1\ldots 6}$ that 
are automatically taken into account by the redefinition
of the Wilson coefficients in Eq.~(\ref{eff}).}
\label{bsll01}
\end{figure}

$\bullet \,$ The effective coefficients $\widetilde{C}_{7-10}^{\rm eff}$ 
appearing in Eq.~(\ref{effmod}) are defined as:
\bea
\widetilde{C}_7^{\rm eff} &=& \frac{4 \pi}{\alpha_s(\mu)} C_7(\mu)
-\frac{1}{3} C_3(\mu) -\frac{4}{9} C_4(\mu)
-\frac{20}{3} C_5(\mu) -\frac{80}{9} C_6(\mu)\nonumber \\
\widetilde{C}_8^{\rm eff} &=& \frac{4 \pi}{\alpha_s(\mu)} C_8(\mu)
 + C_3(\mu) -\frac{1}{6} C_4(\mu)
+ 20  C_5(\mu) -\frac{10}{3} C_6(\mu) \nonumber\\
\widetilde{C}_9^{\rm eff}(s) &=& 
  \frac{4 \pi}{\alpha_s(\mu)} C_9(\mu)
+ \sum_{i=1}^6 C_i(\mu) \gamma^{(0)}_{i9}  \ln\left(\frac{m_b}{\mu}\right)
\nonumber\\ &+& h\left(z,s \right) \left( 
\frac{4}{3} C_1(\mu) + C_2(\mu)   
+ 6 C_3(\mu) + 60 C_5(\mu) \right)
\nonumber\\ &+& h(1,s) \left(  
-\frac{7}{2} C_3(\mu)-\frac{2}{3} C_4(\mu)-38 C_5(\mu)-\frac{32}{3} C_6(\mu) \right)
\nonumber\\ &+& h(0,s) \left(  
-\frac{1}{2} C_3(\mu)-\frac{2}{3} C_4(\mu)- 8 C_5(\mu)-\frac{32}{3} C_6(\mu) \right)
\nonumber\\ &+& 
\frac{4}{3} C_3(\mu)+ \frac{64}{9} C_5(\mu)+ \frac{64}{27} C_6(\mu)\nonumber\\
\widetilde{C}_{10}^{\rm eff} &=& 
 \frac{4 \pi}{\alpha_s(\mu)}   C_{10}(\mu)~,
\label{eff}
\eea
where 
\bea
h(z,s) &=&  - \frac{4}{9} \ln(z) + \frac{8}{27} + \frac{16}{9}\frac{z}{s} 
              - \frac{2}{9} \left( 2+\frac{4\, z}{s} \right)
              \sqrt{\left|\frac{4\,z-s}{s}\right|} \times \no\\
       &\times& \left\{ \ba{ll}
 2 \arctan \sqrt{\frac{s}{4\,z-s}} & \mbox{for} \, s < 4\,z\, ,   \\ 
 \ln \left(\frac{\sqrt{s} + \sqrt{s - 4\,z}}{\sqrt{s} - 
\sqrt{s - 4\,z}} \right) -i\,\pi \qquad &\mbox{for}\,  s > 4\,z  \, . \ea \right.
\eea
Note that specific one- and two-loop and matrix-element contributions  
of the four-quark operators ${\cal O}_{1-6}$ 
(including the corresponding bremsstrahlung contributions) such as the one 
shown in Fig.~\ref{bsll01} are automatically  included by 
the redefinition of the 
Wilson coefficients $C_7$, $C_9$ and $C_{10}$
given in (\ref{eff}). In fact, using this redefinition,    
the bremsstrahlung and virtual corrections that were
shown in Fig.~\ref{bsll03} automatically take  these effects into 
account. The Wilson coefficients $C_i$ in (\ref{eff}), 
which are needed to NNLL precision, 
are given in \cite{MisiakBobeth}.

\frenchspacing
\footnotesize
\begin{multicols}{2}

\end{multicols}
\end{document}